\documentclass{optica-article}

\journal{opticajournal} % for journals or Optica Open

\articletype{Research Article}

\usepackage{lineno}
\begin{document}

\title{Quantum Vacuum in Matter}

\author{Andrey Baydin,\authormark{1,2,*} Hanyu Zhu,\authormark{1,2,3,4,\dag}, Motoaki Bamba\authormark{5}, Kaden Hazzard\authormark{2,4} and Junichiro Kono\authormark{1,2,3,4,\ddag}}

\address{
\authormark{1}Department of Electrical and Computer Engineering, Rice University, 6100 Main Street, Houston, Texas 77005, USA\\
\authormark{2}Smalley--Curl Institute, Rice University, 6100 Main Street, Houston, Texas 77005 USA\\
\authormark{3}Department of Materials Science and NanoEngineering, Rice University, 6100 Main Street, Houston, Texas 77005, USA\\
\authormark{4}Department of Physics and Astronomy, Rice University, 6100 Main Street, Houston, Texas 77005, USA\\
\authormark{5}Department of Physics, Yokohama National University, Yokohama, Kanagawa, 240-8501, Japan\\
}

\email{\authormark{*}baydin@rice.edu}
\email{\authormark{\dag}hanyu.zhu@rice.edu}
\email{\authormark{\ddag}kono@rice.edu}

% use {asbstract*} to suppress the copyright line. Copyright information will be added in production

\begin{abstract*} 
An intriguing consequence of quantum field theory is that vacuum is not empty space; it is full of quantum fluctuating electromagnetic fields, or virtual photons, corresponding to their zero-point energy, even though the average number of photons is zero. These short-lived vacuum fluctuations are behind some of the most fascinating physical processes in the universe, including spontaneous emission, the Lamb shift, and the Casimir force.
Recent theory and experiments indicate that the properties of materials placed in photonic cavities may be altered, even in the complete absence of any external fields, through interaction with the fluctuating vacuum electromagnetic fields. Judicious engineering of the quantum vacuum surrounding the matter inside a cavity can lead to significant and nonintuitive modifications of electronic and vibrational states, producing a ``vacuum dressed'' material. These exciting new ideas have stimulated discussions regarding the fundamental physics of vacuum--matter interactions and also broadened the scope of potential applications using zero-point fluctuations to engineer materials. This Perspective will first discuss recent experimental and theoretical developments on vacuum-modified condensed matter systems, which usually require the realization of the so-called ultrastrong light--matter coupling regime. Then, we will overview some of the most promising cavity designs for enhancing vacuum electromagnetic fields in materials with various energy scales. Finally, we will discuss urgent open questions and technical challenges to be solved in this emerging field.
\end{abstract*}

%%%%%%%%%%%%%%%%%%%%%%%%%%  body  %%%%%%%%%%%%%%%%%%%%%%%%%%
\section{Introduction}
The quantum vacuum is a source of a variety of nonintuitive physical phenomena, including spontaneous emission, the Lamb shift, and the Casimir force~\cite{Milonni1994}. Recent advances in the optical studies of condensed matter have led to the emergence of vacuum-related phenomena that have conventionally been studied in the realm of quantum optics. These studies have not only deepened our understanding of light--matter interactions but have also introduced aspects of many-body correlations inherent in optical processes to condensed matter. In particular, various solid-state cavity quantum electrodynamics (QED) systems have recently emerged, opening exciting opportunities for studying the ultrastrong coupling of quantized light with quantum many-body systems, with long-term implications for quantum information technology. 

\begin{figure}[htp]
    \centering
    \includegraphics[width=0.72\textwidth]{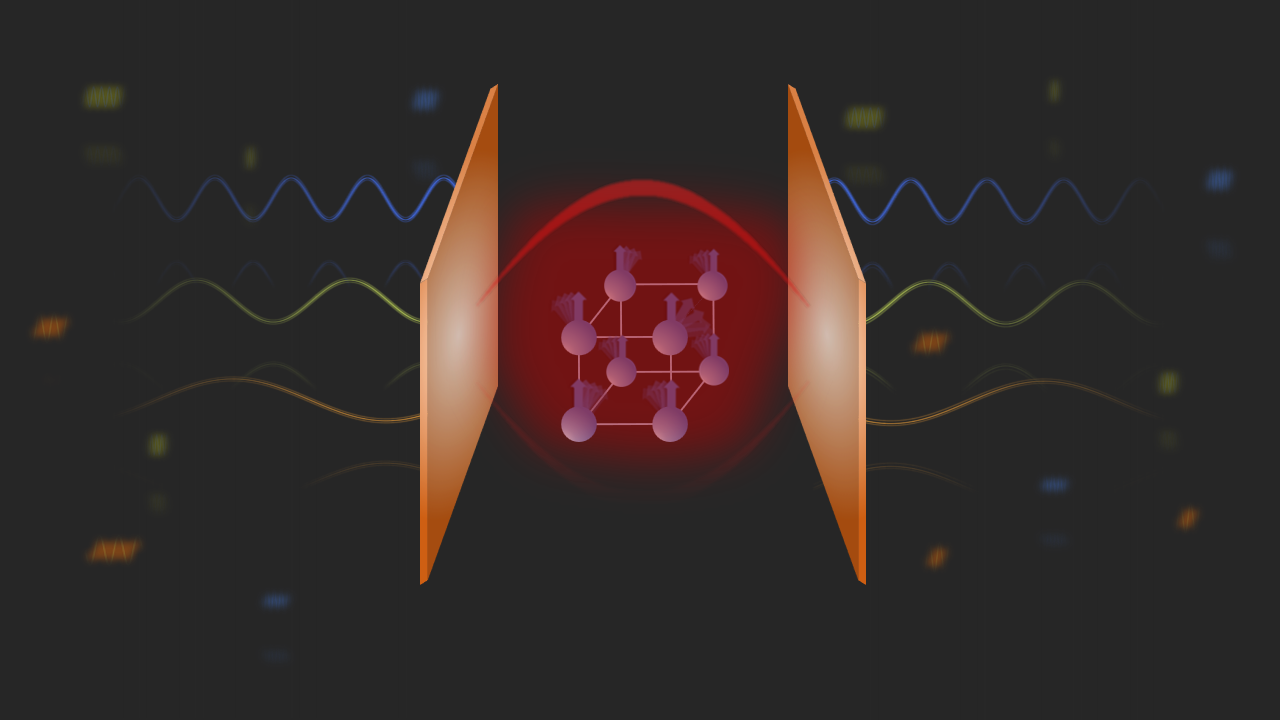}
    \caption{
    An illustration of matter placed between mirrors that form an optical cavity and therefore modify the electromagnetic density of states experienced by matter. This poses a big question: \emph{what is the ground state of matter modified by the cavity vacuum fields?}
    }
    \label{fig:concept}
\end{figure}

During the last decade, robust effects of the quantum vacuum have been experimentally observed in condensed matter cavity QED systems in the ultrastrong coupling (USC) regime~\cite{Forn-DiazEtAl2019RMP,FriskKockumEtAl2019NRP}, including vacuum Rabi splittings comparable to the bare frequencies and vacuum Bloch--Siegert shifts evidencing the breakdown of the rotating-wave approximation (RWA)~\cite{LiEtAl2018NP}. An exciting aspect of such light--matter hybrids is that the ``light field'' that the matter strongly couples with is not an external laser field but the vacuum fluctuation field in the cavity; see a schematic illustration of matter experiencing the vacuum field modified by the cavity in Fig.~\ref{fig:concept}. 

Beyond simply modifying the energy spectra~\cite{Forn-DiazEtAl2019RMP,FriskKockumEtAl2019NRP}, USC has been theoretically predicted to introduce nonclassical properties in the vacuum-dressed eigenstates, and particularly interestingly, in the ground state. The newly produced ground state, represented by a matter--vacuum entangled wavefunction~\cite{CiutietAl05PRB,AshhabNori10PRA}, has characteristics that differ from both the original matter (such as a change from being insulating to metallic) and the ordinary vacuum (such as the addition of the finite number of photons, which are called virtual photons). Examples of nontrivial effects include intrinsic quantum squeezing in the ground states~\cite{CiutietAl05PRB,DeLiberato2017NC}, quantum vacuum radiation from light--matter hybrids~\cite{LiberatoEtAl2007PRL,Hagenmuller2016PRB}, and the Dicke--Hepp--Lieb superradiant phase transition (SRPT; also known as photon condensation)~\cite{HeppLieb1973AoP}.

In addition to two-level systems and band insulators, there have been proposals for observing vacuum effects in superconductors, ferroelectrics/ferromagnets, topological materials, and strongly correlated materials~\cite{PeracaEtAl2020SaS}. Due to the difficulty in accounting all field-coupled terms of the many-body physics in condensed matter, these predicted phenomena are derived under various approximations and not without controversy, both qualitatively regarding the robustness of the theoretical approximations and quantitatively regarding the expectation and interpretation of effects in realistic experimental conditions and methodologies. Therefore, both theoretical and experimental research are necessary to understand and control the vacuum--matter hybrids.  

While there have been a few review articles published on the topic of USC and materials in cavities~\cite{Forn-DiazEtAl2019RMP,FriskKockumEtAl2019NRP,SchlawinEtAl2022APR,LuEtAl2025AOPA}, this Perspective aims to provide the present status of the field and insights for the future. First, we will survey the latest experimental observations related to the impact of cavity vacuum fields on material properties and the theoretical background to contextualize them. We will highlight the recent breakthroughs in increasing the light--matter coupling strength and the consequent emergence of new phenomena. Additionally, we will discuss novel cavity designs for vacuum field engineering and promising cavity--matter systems. We will also present the latest theoretical advancements in condensed matter cavity QED. Finally, we will address current challenges and open questions, such as the effects of the $A^2$ term and gauge invariance, antiresonant versus resonant effects, how long-range interaction by vacuum photons affect correlated electron-spin-lattice systems, and phenomena induced by chiral cavities. We conclude with an outlook for the advent of quantum vacuum electronics.

\section{Theory and Experimental Demonstrations of Cavity--Materials Interactions
\label{theory-experiment}}

\def\oHH{\hat{\mathcal{H}}}
\def\oa{\hat{a}}
\def\oad{\hat{a}^{\dagger}}
\def\ob{\hat{b}}
\def\obd{\hat{b}^{\dagger}}
\def\osigma{\hat{\sigma}}
\def\oS{\hat{S}}
\def\wcav{\omega_{\text{cav}}}
\def\wmat{\omega_{\text{mat}}}
\def\wz{\omega_0}
\def\cs{g}
\def\csz{g_0}
\def\cAA{D}
This section introduces the theoretical basis for a plethora of vacuum-enabled phenomena and then discusses experimental work studying these. We will describe the fundamental models governing the interaction of a single quantized electromagnetic mode with discrete excitations of matter, such as spins -- the quantum Rabi,  Dicke, and  Hopfield models -- including the role of the  $A^2$ term, giving an energy cost for the field that is due to the presence of matter. 

The essential features of the quantum vacuum of matter and photons in a cavity are captured by approximating a realistic system with the simple model  Hamiltonian 
\begin{equation} \label{eq:Dicke}
    \oHH = \hbar\wcav\oad\oa + \frac{\hbar\wmat}{2}\sum_{j=1}^N\osigma_{jz} + \hbar\csz(\oad+\oa)\sum_{j=1}^N\osigma_{jx} + \hbar\cAA(\oad+\oa)^2.
\end{equation}
This is accurate when there is a single relevant cavity mode with a resonance frequency $\wcav$, where $\oa$ ($\oad$) is the annihilation (creation) operator of a photon, and when the matter excitations are described by an ensemble of noninteracting identical two-level systems with a resonance frequency $\wmat$, where $\sigma_{j\xi}$ with $\xi \in \{x,y,z\}$ are the Pauli operators representing the $j$th system ($\osigma_{jx} = \osigma_{j+} + \osigma_{j-}$). The third term represents the light--matter interaction. This model also assumes that the photons interact with all the two-level systems uniformly with coupling strength $\csz$ 
per atom, i.e., that the electromagnetic field profile is uniform across the sample, often a reasonable approximation when many atoms are contained in the mode volume. 

Dicke cooperativity, i.e., the increasing of the matrix element to the collective excitation of the matter, enhances the effective coupling strength between the cavity mode and atomic ensemble as the number of atoms $N$ grows as $\cs = \sqrt{N}\csz$. The coupling strength $\cs$ can be roughly evaluated by measuring the vacuum Rabi splitting ($\sim2\cs$)~\cite{Forn-DiazEtAl2019RMP,FriskKockumEtAl2019NRP}. A normalized coupling strength is defined as $\eta = \cs/\wcav$ or $\eta = \cs/\wmat$, and one defines the USC and deep-strong coupling (DSC) regimes by $\eta > 0.1$ and $\eta > 1$, respectively, where the counter-rotating terms cannot be ignored. The last term in Eq.~\eqref{eq:Dicke} is the $A^2$ term, which originates from expanding the kinetic energy in the particles' kinetic energy of the minimal-coupling Hamiltonian. Its coefficient is bounded by $D>N\csz{}^2/\wmat=\cs^2/\wmat$ from the minimal-coupling Hamiltonian, for which the SRPT is prohibited, as elaborated below~\cite{Rzazewski1975}.  

The model in Eq.~\eqref{eq:Dicke} is called the quantum Rabi model and the Dicke model for $N=1$ and $N>1$, respectively. In the limit of $N\gg1$, the Holstein--Primakoff transformation can be used to provide an accurate approximation of the Dicke model known as the Hopfield model, 
\begin{equation} \label{eq:Hopfield}
    \oHH_{\text{Hopfield}} = \hbar\wcav\oad\oa + \hbar\wmat\obd\ob + \sqrt{N}\hbar\csz(\oad+\oa)(\obd+\ob) + \hbar\cAA(\oad+\oa)^2,
\end{equation}
where $\ob$ ($\obd$) is the bosonic annihilation (creation) operator of a collective excitation in the atomic ensemble.  
In Eqs.~\eqref{eq:Dicke} and \eqref{eq:Hopfield}, the terms such as $\oa\osigma_{j-}$, $\oa\ob$, $\oa\oa$, and their Hermitian conjugates are called the counter-rotating terms. For $\eta\ll1$, these terms can be neglected (the RWA), and the total number of photons and matter excitations $N_{\text{exc}}=\oa^\dagger \oa + (1/2)\sum_j (\osigma_{jz} +1)$ is conserved, and there is correspondingly an emergent $U(1)$ symmetry; in this limit, Eq.~\eqref{eq:Dicke} is known as the Jaynes--Cummings ($N=1$) or Tavis--Cummings ($N>1$) model. However, the RWA fails in the USC and DSC regimes, and the symmetry is reduced to ${\mathbb Z}_2$ associated with simultaneously reflecting $\oa$ and $\osigma_{jx}$ (in the Dicke model) or $\ob$ (in the Hopfield model) simultaneously.

The counter-rotating light--matter coupling terms introduce new physics, such as two-mode quantum squeezing in the ground states. This equilibrium quantum squeezing is particularly interesting as a robust quantum resource with a simple kind of ``passive quantum error correction'': after a decoherence event, the system will thermally relax to the entangled ground state. Also, the ground state of the USC system contains virtual photons. While there are many proposals to convert virtual photons to real ones~\cite{LiberatoEtAl2007PRL,StassiEtAl2013PRL,Hagenmuller2016PRB,CirioEtAl2016PRL}, e.g., quantum vacuum radiation through the Moore–Fulling–Davies effect (also known as the dynamic Casimir effect)~\cite{Moore1970JMP,1976PRSLA}, the challenge of its observation is still ongoing. 

As a step to observing these effects, the first clear evidence of RWA breakdown (influence of the counter-rotating coupling) was observed as the vacuum Bloch--Siegert shift in the Landau polariton system in 2018~\cite{Li2018}. Although the observation of equilibrium quantum squeezing is still ongoing, theoretically, large quantum squeezing is expected at an SRPT critical point~\cite{Hayashida2023}.

While the SRPT was shown to occur in the Dicke model in the absence of the $A^2$ term by Hepp and Lieb in 1973~\cite{Hepp1973}, a no-go theorem accounting for the $A^2$ term in some scenarios appeared just two years later~\cite{Rzazewski1975}. Since then, an ongoing discussion about the viability of the SRPT in light--matter coupled systems has continued. A recent trend is to focus on the Zeeman interaction $\mathfrak{g}\mu_\text{B}\hat{B}\sum_{j=1}^N\hat{s}_{j,z}$ between an ensemble of spins $\{\hat{s}_{j,z}\}$ and the magnetic flux density $\hat{B}$, i.e., spatially varying vector potential~\cite{Manzanares2022}, where $\mathfrak{g}$ is the Land\'{e} $g$-factor and $\mu_\text{B}$ is the Bohr magneton, which may have the potential to cause an SRPT~\cite{Rouse2023} and a Landau polariton system with Rashba spin--orbit coupling was proposed theoretically in 2022~\cite{Manzanares2022}. However, it is still unresolved whether, starting from a more fundamental theory,  quadratic terms (such as the $A^2$ term) may prohibit the SRPT. Experimentally, whereas a nonequilibrium SRPT has been observed in driven cold atoms in 2010~\cite{Baumann2010}, it is quite recent that an equilibrium SRPT was evidently confirmed as a magnonic version in a magnetic material $\mathrm{ErFeO_3}$ by ultrabroadband spectroscopy~\cite{KimEtAl2024}, which evades the no-go theorem by using a collective magnetic mode in place of a real physical cavity. 

The coupling between light and matter in photonic cavities is tunable and has thus inspired numerous theoretical proposals to engineer materials properties by coupling with electromagnetic quantum fluctuations~\cite{SchlawinEtAl2022APR}. Here, the simple Dicke and related models may often fail to capture the mechanisms by which the materials' properties are modified, as either the matter degrees of freedom do not reduce to simple two- (or few-) level systems, many light modes are relevant, or both; we provide examples later. Moreover, in many analytical and numerical approaches, the cavity--matter coupling constant was treated as a free parameter instead of being derived from first principles or from independent experimental measurements, and the photonic modes were often truncated~\cite{schafer_making_2021}. Meanwhile, some theoretical studies suggested that the inclusion of the full spectrum and the energy stored by the cavity-forming materials may both be important for the shift of ground-state energy of the material placed inside the cavity~\cite{saez-blazquez_can_2023}. A more complete theoretical model would require treating the two materials coupled by long-range electromagnetic interactions, which has not yet been calculated using any first-principles approach. 

\begin{figure}[htp]
    \centering
    \includegraphics[width=\textwidth]{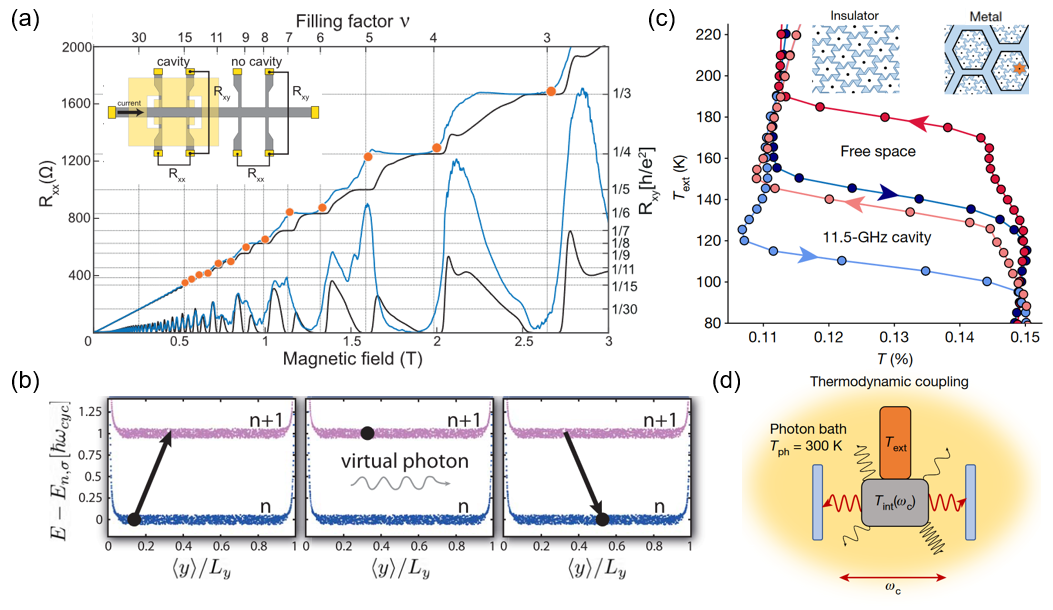}
    \caption{Recent experimental demonstration of cavity-induced phenomena. 
    (a)~Cavity-induced finite longitudinal resistance in a quantum Hall system at integer filling factors. (b)~Schematic of a localized electron at an even-order Landau level scattered into another localized state by absorbing and emitting a virtual photon. Adapted from Ref.~\cite{AppuglieseEtAl2022S}. (c)~Cavity-induced apparent change of temperature for the metal--insulator phase transition in 1T-TaS$_2$. (d)~Possible cavity-mediated radiative heat exchange between an externally cooled sample and the environmental photon bath. Adapted from Ref.~\cite{JarcEtAl2023N}.}
    \label{fig:exp}
\end{figure}

Despite the theoretical challenges, there are already several experimental reports presenting evidence that the coupling strength can be sufficiently strong to affect material properties in cavities~\cite{AppuglieseEtAl2022S,JarcEtAl2023N,KumarEtAl2024JACS,ThomasEtAl2025TJoCP,EnknerEtAl2025N}. 
One example is a cavity-modified quantum Hall system, where a breakdown of the topological protection of the integer quantum Hall effect was observed~\cite{AppuglieseEtAl2022S}. This observation is rather striking because the quantum Hall effect is usually robust and topologically protected against weak local perturbations. More specifically, in this experiment, a two-dimensional electron gas in a GaAs quantum well was placed in a deep subwavelength millimeter-wave resonator. For odd-integer filling factors $\nu=5-11$, the longitudinal resistivity became comparable with the midgap values that are typically obtained when there is no quantum Hall state, and the Hall resistivity shifted towards the quantized value of the next filling factor $\nu-1$, as shown in Fig.~\ref{fig:exp}(a). The authors interpreted their observations in terms of modification of the effective g-factor by the gradient of fluctuation fields~\cite{AppuglieseEtAl2022S,ArwasCiuti2023PRB}. For even-integer filling, the change in resistivity is much smaller and was explained as a result of long-range cavity-mediated electron hopping, illustrated in Fig.~\ref{fig:exp}(b). In this picture, a localized electron at the $n$-th Landau level is scattered into another localized state by absorbing and emitting a virtual photon. Later, a theoretical follow-up study demonstrated that this weakening of topological protection is due to the existence of the lower polariton mode, which is softer than the Kohn mode~\cite{RokajEtAl2023PRL}. Most recent experimental work points out the importance of strong vacuum electric field gradients in cavity quantum Hall systems~\cite{EnknerEtAl2025N}. 

Another recent example of cavity-induced modifications of materials properties is the observation of a cavity-modified charge density wave phase transition in the material 1T-$\mathrm{TaS_2}$ [Fig.~\ref{fig:exp})(c)] when the resonance frequency of the cavity is much lower than the frequencies of the common oscillators, including plasmons, infrared-active phonons, or band-to-band transitions~\cite{JarcEtAl2023N}. The apparent transition temperature decreased monotonically from the free-space value as the cavity volume increased. One possible explanation is that, rather than coherently coupling with the cavity, 1T-$\mathrm{TaS_2}$ may have exhibited an anomaly in emissivity at the GHz frequencies and then exchanged energy with the room-temperature environment more efficiently through a close-to-resonance cavity, schematically shown in Fig.~\ref{fig:exp}(d). Such a ``thermal Purcell effect''~\cite{fassioli_controlling_2024} does not require strong coupling and could be a contributing factor to many other cavity engineering schemes, such as vibrational polaritonic chemistry~\cite{SidlerEtAl2021JPCL}. Interestingly, we note that if thermal equilibration in materials is slow with respect to the physical or chemical process under consideration, the cavity can serve as a selective heater or cooler of a subsystem (electron, lattice, or spin) to trigger a preferred property without negatively disrupting the entire material. 

Strictly speaking, no experiments so far have unambiguously proved vacuum engineering of the ground state itself, even if the materials reach the collective USC regime ($\sqrt{N}$ enhancement of $\cs$). Nevertheless, the aforementioned results clearly demonstrated that pathways and effects unexpected in the original theoretical predictions are vital for cavity QED engineering~\cite{bartolo_vacuum-dressed_2018} and shall motivate more extensive searches in the future. 

\section{Cavity Design and Materials for Ultrastrong and Deep-Strong Coupling}
As evident from the previous section, the USC and DSC regimes are paramount in vacuum material engineering, and therefore, here, we will review and highlight the types of optical cavities and their suitability for particular materials systems. 

USC and DSC can be achieved via engineering an electromagnetic environment using an optical cavity. However, the coupling strength has to be optimized to observe vacuum-induced phenomena in cavities. The coupling strength per atom $g_0$ can be written (for systems where the two-level approximation is valid) as 
\begin{align}
    \csz = \frac{\mu_{12}E_\text{vac}}{\hbar}, 
    \label{eq:coupling}
\end{align}
where $\mu_{12}$ is the dipole moment of the resonant optical transition $|1\rangle \rightarrow |2\rangle$, $\hbar$ is the reduced Planck constant, and $E_\text{vac}$ is the strength of the vacuum fluctuation field in the cavity given by 
\begin{align}
    E_\text{vac} = \sqrt{\frac{\hbar\omega_0}{2\varepsilon_\text{r}\varepsilon_0 V_\text{m}}}.
    \label{eq:vac-field}
\end{align}
Here, $\varepsilon_\text{r}$ is the dielectric constant of the medium inside the cavity, $\varepsilon_0$ is the vacuum permittivity, and $V_\text{m}$ is the mode volume. Thus, choosing materials with large dipole moments and maximizing the vacuum fluctuation field strength by cavity design can lead to the USC and DSC regimes. 

The most common types of cavities used in USC and DSC studies are diffraction-limited optical cavities, such as Fabry-P\'{e}rot (FP) cavities~\cite{BaranovEtAl2020NC,MavronaEtAl2021AP} and photonic-crystal cavities (PCC)~\cite{ZhangEtAl2016NP,LiEtAl2018NP}, and subdiffraction cavities such as plasmonic resonators~\cite{ScalariEtAl2012S,BayerEtAl2017NL}. The former tends to have high $Q$-factors but weak field enhancement; the latter can have strong field enhancement but lower $Q$-factors. The cavity choice is usually driven by the sample, cavity fabrication requirements, and measurement tools. 

\subsection*{Diffraction limited optical cavities}
The simplest cavity structure that is usually considered is the FP cavity, which confines light between two mirrors or interfaces~\cite{BaranovEtAl2020NC}. When two media with different refractive indices and thicknesses are stacked periodically, they form a one-dimensional (1D) PCC, which has allowed and forbidden bands (or band gaps) for photons. If the periodicity is broken by introducing a ``defect'' layer, such as a thicker layer, a photonic mode is formed inside a photonic band gap. When the photonic mode is close to the center of the photonic band gap, the photonic crystal on both sides of the defect layer behaves as two parallel mirrors, trapping photons within the defect layer like a cavity. The localized mode leads to an enhancement of the electric field strength within a finite region. USC can be achieved by placing a material with a large electric dipole moment in a region with a high electric field. As the formation of a PCC exploits the concept of constructive and destructive interference, the smallest cavity mode volume that can be achieved in the PCC is limited by diffraction to $\sim(\lambda/2)^3$.

Recently, a three-dimensional (3D) PCC has been realized, which has considerable improvements over 1D-PCCs~\cite{TayEtAl2023}. A 3D-PCC is formed by stacking a series of silicon rods stacked with alternating orthogonal orientations in a ``woodpile'' structure~\cite{LiuEtAl2007OEO,ChenEtAl2008JAP,Hagenmuller2016PRB}. As the electromagnetic field is confined in all directions in the 3D-PCC, the effective mode volume, $V_\text{m}$, of the woodpile cavity is much smaller than that of the corresponding 1D-PCC, providing stronger vacuum--matter coupling. Thus, Tay \textit{et al.} demonstrated multimode USC between multiple photonic modes of a terahertz 3D-PCC and a Landau-quantized two-dimensional electron gas (2DEG) in GaAs. They identified two distinct multimode light-matter coupling scenarios in a 3D-PCC device. Unlike conventional FP cavities and 1D PCCs, the discrete in-plane translational symmetry of the 3D-PCCs allows the cavity photons to couple with matter excitations with a set of finite in-plane reciprocal lattice vectors. Although the dipole approximation remains valid (because the spatial profiles of the cavity modes are nearly uniform on the scale of the electron cyclotron orbits), the spatial variation of the cavity field significantly affects the mixing between cavity modes via the CR. This results in strong correlations between vacuum electromagnetic fields at different frequencies~\cite{TayEtAl2023}. 

Another improvement over the 1D-PCCs is a Tamm cavity. It can be formed by cutting a 1D-PCC in half and depositing a thin metal layer (i.e., Au), as shown in Fig.~\ref{fig:cavities}(b). In contrast to the 1D-PCC, which localizes the electric field at the two boundaries of the defect layer, the electric field of the Tamm mode is localized at only one boundary of the layer coated with the metal layer. Thus, the local electric field in the Tamm cavity is higher than that in the corresponding 1D-PCC, leading to a higher coupling strength. The polarization and effective mode volume of the Tamm mode can be further manipulated by patterning the metal layer~\cite{MesselotEtAl2020AP}. Employing a metal-stripe pattern or a rectangular lattice design for the metal layer is expected to alter the field distribution of the Tamm mode, reducing the mode volume and creating in-plane anisotropy.

\subsection*{Sub-diffraction limited cavities}

The cavity mode volume, $V_\text{m}$, is the defining parameter for the light--matter coupling strength ($g\propto1/\sqrt{V_\text{m}}$) in a cavity QED system~\cite{HagenmullerEtAl2010PRB,ScalariEtAl2012S,RajabaliEtAl2021NP}; see Eqs.\,\eqref{eq:coupling} and \eqref{eq:vac-field}. In many cases, the ground-state modification by cavities relies on non-resonant coupling terms, so a stronger single-particle coupling strength, i.e., larger dipole moment in smaller cavities, tends to be more beneficial than higher quality factors of the resonances or cavities. 
Combining materials with positive and negative permittivity, one can confine a resonant mode with very large momenta, enabling unprecedented coupling strengths of light--matter interaction~\cite{Forn-DiazEtAl2019RMP,FriskKockumEtAl2019NRP,BayerEtAl2017NL}. Planar THz metamaterials consist of arrays of $LC$ resonators where the electric field confinement is achieved in a few micrometer capacitance gaps. The DSC regime can be realized by coupling a single photonic mode in such metamaterials with a single matter mode of low-frequency excitations such as cyclotron resonance in a 2DEG~\cite{BayerEtAl2017NL} or soft phonon modes~\cite{Baydin2025Arxiv}. With mode-multiplexing approaching, the normalized coupling strength can even become $>3$, which gives vacuum ground state populations of virtual photons exceeding 1~\cite{MornhinwegEtAl2024NC}. Terahertz fields can be further enhanced in metallic nanoslits or nanogaps even if the thickness and slit/gap size are smaller than the skin-depth of the material~\cite{SeoEtAl2009NP}, resulting in USC in systems with smaller oscillator strengths than CR or soft phonons as mentioned above~\cite{kim_cavity-mediated_2024}.

At optical frequencies, metallic nanostructures can support localized surface plasmons to enhance single-particle light-matter coupling. For example, the usually weak spontaneous Raman scattering, which scales as the fourth power of the cavity field strength, can be strong enough to create USC between photons and phonons for a single molecule between a gold substrate and the tip of a gold nanoparticle in a ``picocavity'' [Fig.~\ref{fig:cavities}(c)]. The effective $V_\text{m}$ was assumed to be $ 10^{-2} \, \text{nm}^3$, which means that the fluctuating electric field would exceed 100\,MV/cm, much larger than the damaging threshold of the molecule if the photon is real on any relevant timescale~\cite{benz_single-molecule_2016}. One would thus expect that the electronic energy levels should be modified, which can also be understood from another viewpoint: the electromagnetic fluctuation is essentially the van der Waals--Casimir force between the tip and molecule, known to broaden and/or shift the molecular energy levels in scanning tunneling microscopy, except that the spectral function is altered by nonlocal cavity resonance. However, in this work, the electronic states were not measured and no obvious vibrational frequency shift was found in the data as a result of changes in the atomic potential landscape. 

Another way to achieve extreme confinement for light from the visible band to the far-infrared band is through hyperbolic materials, especially layered compounds~\cite{ashida_cavity_2023}. Anisotropic plasmons can be found in semiconductors like WSe$_2$ and semimetals like graphene, which offer an electronically tunable plasmon response as shown in Fig.~\ref{fig:cavities}(d)~\cite{epstein_far-field_2020, sternbach_programmable_2021}, as well as metals~\cite{ruta_good_2024}. Anisotropic phonon-polaritons were demonstrated in insulators and generally exhibit lower loss, higher quality factors, as well as multiple modes [see Fig.~\ref{fig:cavities}(e)]~\cite{dubrovkin_resonant_2020,xu_phonon_2023,herzig_sheinfux_high-quality_2024}. In theory, there is no lower bound on the nominal cavity volume, but practically the fluctuating field is limited by two factors: the roughness and damage of the materials from the fabrication process, and the fraction of electromagnetic energy in the hybrid mode. An ultimate field strength useful for materials engineering may be estimated as that of a charged harmonic oscillator in a unit cell placed at a distance of the van der Waals gap, i.e., on the order of 1\,MV/cm.

\subsection*{Chiral cavities}
\begin{figure}
    \centering
    \includegraphics[width=0.95\textwidth]{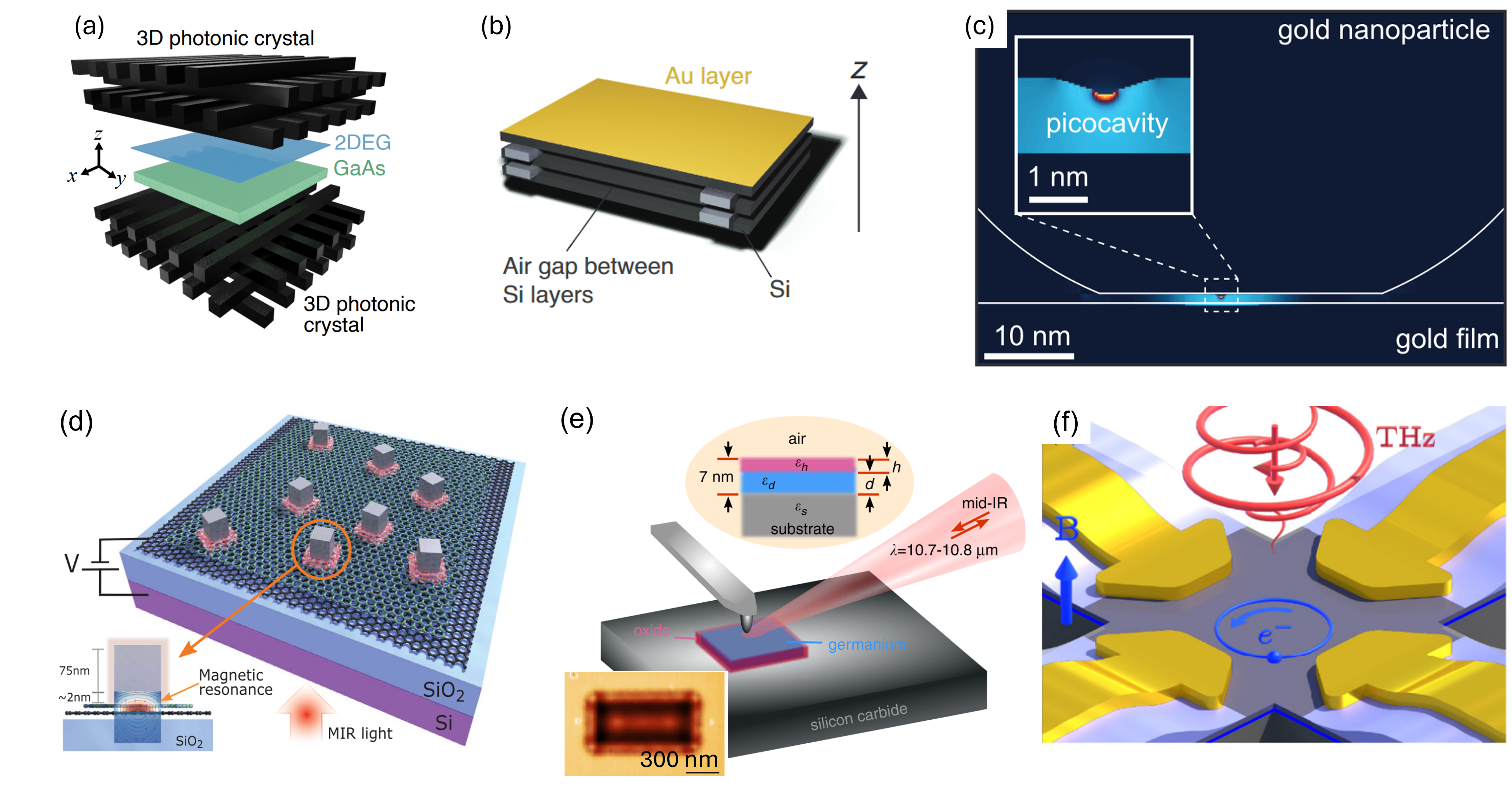}
    \caption{Examples of novel cavities to engineer light-matter coupling.
    (a) 3D-PCC. Adapted from Ref.~\cite{TayEtAl2023}
    (b) Tamm cavity. Adapted from Ref.~\cite{MesselotEtAl2020AP}
    (c) ``Picocavity" formed by the gold substrate and a tip on a gold nanoparticle with a molecule trapped nearby, enhancing Raman scattering to the strong coupling regime. Adapted from Ref.~\cite{benz_single-molecule_2016}
    (d) The nanoplasmonic cavity tunable in the mid-infrared frequencies formed by silver nanocubes and graphene. Adapted from Ref.~\cite{epstein_far-field_2020} 
    (e) SiC phonon-polaritonic nanocavity with an apparent quality factor above 100. Adapted from Ref.~\cite{dubrovkin_resonant_2020}.
    (f) The linearly polarized nanoantenna modes are coupled to the time-reversal symmetry-breaking inter-Landau-level transition of the 2DEG (in blue), giving rise to nondegenerate circularly polarized modes. Adapted from Ref.~\cite{AndbergerEtAl2023}.
    }
    \label{fig:cavities}
\end{figure}

A so-called chiral cavity can confine electromagnetic modes of specific chirality, i.e., circular polarization, which can have a profound impact on the nature of the ground state of a vacuum--matter hybrid in the USC regime. For example, if only right- or left-circular polarized light modes are supported by the chiral cavity, it will result in the breaking of time-reversal symmetry inside the cavity. Thus, chiral vacuum fields inside a chiral cavity are a promising route to topological band engineering, similar in spirit to the Floquet approach of material control, but without requiring external light~\cite{WangEtAl2019PRB,HubenerEtAl2021NM,DagRokaj2023}. 
To date, there have been several studies of chiral cavities with mirror symmetry breaking, but only recently, a few reports of chiral cavities with time-reversal symmetry breaking have appeared. Su\'arez-Forero \textit{et al}.\ demonstrated a FP type chiral cavity operating in the near-infrared~\cite{Suarez-ForeroEtAl2023} and Andberger \textit{et al.}~\cite{AndbergerEtAl2024PRB} and Aupiais \textit{et al.}~\cite{AupiaisEtAl2024} demonstrated plasmonic chiral cavities for the terahertz frequency range. Figure~\ref{fig:cavities}(b) shows an illustration of the chiral cavity where an array of nanoantennas couples to the time-reversal symmetry-breaking inter-Landau-level transition of the 2DEG. Such a coupling gives rise to nondegenerate circularly polarized modes~\cite{AndbergerEtAl2024PRB}. The authors comment that in order to study the effect of chiral vacuum fluctuations on various samples, they can be brought close to the capacitor gap~\cite{AndbergerEtAl2024PRB}. While plasmonic cavities offer an extremely large electric field enhancement, their $Q$-factors remain low.

For FP-type cavities with time-reversal symmetry breaking, Faraday mirrors have been proposed~\cite{HubenerEtAl2021NM,SunEtAl2022CS}, which can induce a phase difference and result in chiral standing waves of opposite circular polarization peaking at different locations inside the cavity. Tay \textit{et al}.\ have designed a chiral cavity based on a magnetoplasma in InSb. Due to the light effective mass of electrons in InSb, CR occurs at THz frequencies even at small magnetic fields ($B\sim0.2$\,T), thus absorbing CR-active chiral standing waves while waves of the other CR-inactive mode are not affected. The $Q$-factor and mode profile of the resonance in the chiral 1D-PCC are comparable to those in a conventional 1D-PCC. The ellipticity is close to 1 at the position of the maximum electric field and remains uniform in the plane perpendicular to the 1D-PCC structure. Therefore, such a chiral 1D-PCC with time-reversal symmetry breaking is promising for studying the effects of chiral vacuum electromagnetic fields on 2D materials such as graphene, van der Waals materials, and quantum wells~\cite{TayEtAl2024}.

\section{Open Questions, Challenges to Overcome, and Conclusions}
The nascent field of quantum vacuum engineering in strongly coupled polaritonic systems has demonstrated several proof-of-principle milestones, as overviewed above, but fully realizing its promise requires advances in both theory and experiment.

Theoretically, materials coupled with cavities require a rethinking of our understanding of many-body systems and the development of practical tools for calculating properties. Three broad areas are apparent: (1)~incorporation of the extremely long-range interactions induced by light into many-body theory and an understanding of their consequences, (2)~establishment of a set of core models and paradigms for thinking about quantum vacuum-modified materials, and (3)~clarifying the role of entanglement (versus merely strong correlations) in quantum vacuum-modified materials. The last point is a field just at its inception in real solid-state materials more broadly, and vacuum-modified materials can play a special role here with their ability to induce and control entanglement. We discuss each in turn.  

The first point, the challenges and the path forward for incorporating the long-range electromagnetic interactions is best appreciated by first examining the contrast with the typical theory of materials. A key concept underpinning phenomena in materials is \emph{locality}: any point in space cannot instantaneously interact with particles far away. The consequences of locality are numerous and important:  local perturbations perturb locally (LPPL), e.g., in a gapped ground state, a perturbation at a point $\vec{r}$ in space has an effect that decays exponentially in the distance from $\vec{r}$~\cite{bachmann2012automorphic}; the Mermin--Wagner theorem forbids spontaneous symmetry breaking of a continuous symmetry in spatial dimension $d<2$~\cite{mermin1966absence}; correlation clustering, i.e., that gapped systems have correlations between regions that decay exponentially with the distance between the regions~\cite{hastings2006spectral,nachtergaele2006lieb};  the entanglement between two regions in a gapped ground state is proportional to the area of their interface, i.e., satisfies the \emph{Area Law} (proved in 1D in Ref.~\cite{hastings2007area});  and even constrains the admissible types of particle statistics (fermion, boson, anyon)~\cite{doplicher1971local}. 

This notion of locality must be abandoned when there is coupling through a cavity since in just a femtosecond -- a timescale that is comparable to or faster than the time for electrons to tunnel between neighboring lattice sites -- light can transmit information over a micrometer. Therefore, all of the results above (and more) must be updated accordingly.  However, not all hope is lost, as the additional nonlocal interactions still have special features that constrain their effects. The two most important are $k$-locality and the simple mode profile of the interactions. While the interactions are no longer geometrically local, they are still \emph{$k$-local} (for $k=2$), which means that they are of the form $\sum_{ij} C_{ij} {\hat A}_{i} {\hat B}_j $ where ${\hat A}_i$ and ${\hat B}_j$ are operators acting only on finite space--time regions, and $C_{ij}$ is a coefficient matrix arbitrary up to enforcing hermiticity.  Moreover, the structure of the coefficient function $C_{ij}$ is relatively simple: often, it varies slowly compared to microscopic length scales and may factorize, e.g., $C_{ij}=u_i  v_j$. The $k$-locality and simple coefficients strongly constrain the behavior, and an important direction for quantum vacuum engineering is to establish the consequences of having Hamiltonians that are sums of geometrically local and 2-local terms. We note that while the collective light mode of a cavity is the paradigmatic example of this, separation of timescales between a fast mode and other slower degrees of freedom is common, and the associated loss of locality, including systems with multiple magnetic degrees of freedom. 

The second point, the necessity to establish core ideas to understand the complexity of real materials, is another key necessity and opportunity for theory. In ordinary matter, a few key models and ideas explain a wealth of phenomena: models include spin models, noninteracting systems, Hubbard, and Kondo models, while characteristic physical mechanisms include direct and superexchange and their consequences, Landau levels, and band structure characteristics. To progress, theory must answer the question: What are the core concepts with rich explanatory power for quantum vacuum-modified matter? The Dicke model is but one such case -- rather than coupling with independent two-level systems, the cavity can couple with metals, quantum Hall systems, chemical reactions, and more. A suite of key models and results must account for these. The toy models and phenomena, as well as these models and concepts, must be complemented by advancements in numerical techniques.

The final point is driven by the important role that entanglement plays in many quantum vacuum-modified systems; of particular interest is the large amounts of potentially useful entanglement, such as squeezing, that are induced by the quantum vacuum.  However, small imperfections beyond toy models can disturb this fragile entanglement -- in particular, coupling with environmental degrees of freedom can decohere it. It is important to understand what the key mechanisms that can decohere this entanglement are. This is part of a broad emerging field of entanglement in solid-state systems, but one particularly necessitated given the key role entanglement plays in models of quantum vacuum-modified systems. 

Experiments on cavity materials engineering are still in their early stages. Since the USC regime has been achieved in cavities where the electric field is confined to a small volume, it is sometimes challenging to comprehensively probe the effects of USC on macroscopic material properties and/or phase transitions with more conventional direct methods such as transport, heat capacity, or bulk spectroscopy measurements (Section~\ref{theory-experiment}). Instead, micro- or nanoscale spectroscopy is often necessary and may be subject to interpretation and/or artifacts, so it is crucial to systematically establish methodologies to reliably delineate the spectral responses between materials and microcavities, especially deep-subwavelength polaritonic cavities, starting with rigorous control experiments with well-understood systems. For testing the quantum Rabi model in the USC regime, the search for materials with large oscillator strength and cavity designs continues, as it requires the coupling of a single spin/atom with a bosonic field (the coupling strength can no longer be enhanced by $\sqrt{N}$). To incorporate materials with large dielectric functions in cavities, particularly conductive materials such as superconductors, care should be taken in the geometry choice of cavity modes and materials to maximize the coupling strength with reasonable uniformity. Lastly, new design principles should be developed with fabrication feasibility in mind for cavities that can reversibly and continuously break spatial and/or time-reversal symmetry. 

In conclusion, we have provided a perspective on a rapidly emerging and burgeoning field of quantum vacuum in matter. There have been many stimulating theoretical proposals and debates, in particular, surrounding the problem of the superradiant phase transition. Experimentally, the field is still in its early stages, where only a few claims of vacuum modification of matter exist. In the last section, we summarized different approaches to achieving strong coupling of vacuum fluctuations fields with matter in various optical cavities. Overall, research in this field is of high importance and has an impact on the realization of vacuum-enabled electronics.  

\bibliography{2-bibliography}
\end{document}